\documentclass[%
 aip,
 amsmath,amssymb,
 reprint,%
]{revtex4-1}

\usepackage{graphicx}
\usepackage{dcolumn}
\usepackage{bm}

\usepackage[utf8]{inputenc}
\usepackage[T1]{fontenc}
\usepackage{mathptmx}
\usepackage{siunitx}

\begin{document}

\preprint{AIP/123-QED}

\title[IQ Mixer Calibration for Superconducting Circuits]{IQ Mixer Calibration for Superconducting Circuits}

\author{S. W. Jolin}
\email{shanjo@kth.se}
\affiliation{Nanostructure Physics, KTH Royal Institute of Technology, Stockholm, Sweden
}%

\author{R. Borgani}%
\affiliation{Nanostructure Physics, KTH Royal Institute of Technology, Stockholm, Sweden
}%

\author{M. O. Tholén}
\affiliation{Nanostructure Physics, KTH Royal Institute of Technology, Stockholm, Sweden
}%
\affiliation{Intermodulation Products AB, Segersta, Sweden
}%

\author{D. Forchheimer}
\affiliation{Nanostructure Physics, KTH Royal Institute of Technology, Stockholm, Sweden
}%
\affiliation{Intermodulation Products AB, Segersta, Sweden
}%

\author{D. B. Haviland}
\affiliation{Nanostructure Physics, KTH Royal Institute of Technology, Stockholm, Sweden
}%

\date{\today}

\begin{abstract}
An important device for modulation and frequency translation in the field of circuit quantum electrodynamics is the IQ mixer, an analog component for which calibration is necessary to achieve optimal performance. In this paper, we introduce techniques originally developed for wireless communication applications to calibrate upconversion and downconversion mixers.
A Kalman filter together with a controllable carrier frequency offset calibrates both mixers without removing them from the embedding measurement infrastructure. 
These techniques can be embedded into room temperature control electronics and they will find widespread use as circuit QED devices continue to grow in size and complexity.
\end{abstract}

\maketitle

\section{\label{sec:level1} Introduction}

The superconducting qubit processors are continuously growing in scale and complexity, placing increasing demands on the performance of room temperature electronics.
An essential component is the IQ mixer. 
In circuit quantum electrodynamics (QED) \citep{Krantz}, IQ mixers are used to upconvert control signals from classical electronics and downconvert signals emitted from the quantum circuit.
In wireless radiofrequency (RF) and microwave systems, IQ mixers are particularly useful for Quadrature Amplitude Modulation (QAM) schemes, commonly used for cost-efficient and high-data-rate systems \citep{MicrowaveComm, OFDMbook, WirelessSystems, Gu2018, jKim}.
Applications include point-to-point communication \citep{MicrowaveComm} and various radar technologies \citep{TheoryModAntenna, RadarPrinciples}.

IQ mixers are analog components, which by their very nature require calibration.
The error we calibrate for in this paper is the so-called mixer imbalance. 
For upconversion, the imbalance manifests as the generation of a spurious signal called the image, accompanying every desired signal.
For downconversion imbalance prevents unambiguous separation of the signal and image sidebands in a heterodyne detection scheme (or equivalently, a low-IF receiver setup).
In either case, calibration of the IQ mixer allows for conversion to a single sideband of the carrier frequency and suppression of its image, thus leading to more efficient use of frequency space.
This efficiency becomes critical, for example, when implementing high-fidelity qubit gates \cite{Bengtsson2020} for quantum computing and applications where multiple qubits are controlled and measured on the same signal line.
Such frequency-domain multiplexing of signals is also highly desirable in modern communication technology \cite{Windisch2007}.

Technological progress in direct synthesis and sampling of RF signals is rapidly progressing \citep{Andersson2020, kalfus2020}, but until this technology covers the entire frequency band used in circuit QED, IQ mixers are an unfortunate necessity. 
However we can use various digital techniques to compensate for their imperfections.
This is occasionally referred to as the dirty RF paradigm, where imperfect, or dirty, RF devices are improved through digital means \citep{dirtyRF}.
In this paper, we combine ideas originally developed for orthogonal frequency-division multiplexing (OFDM) applications to calibrate mixers in a circuit-QED setting.

\begin{figure}
\includegraphics[scale=0.5]{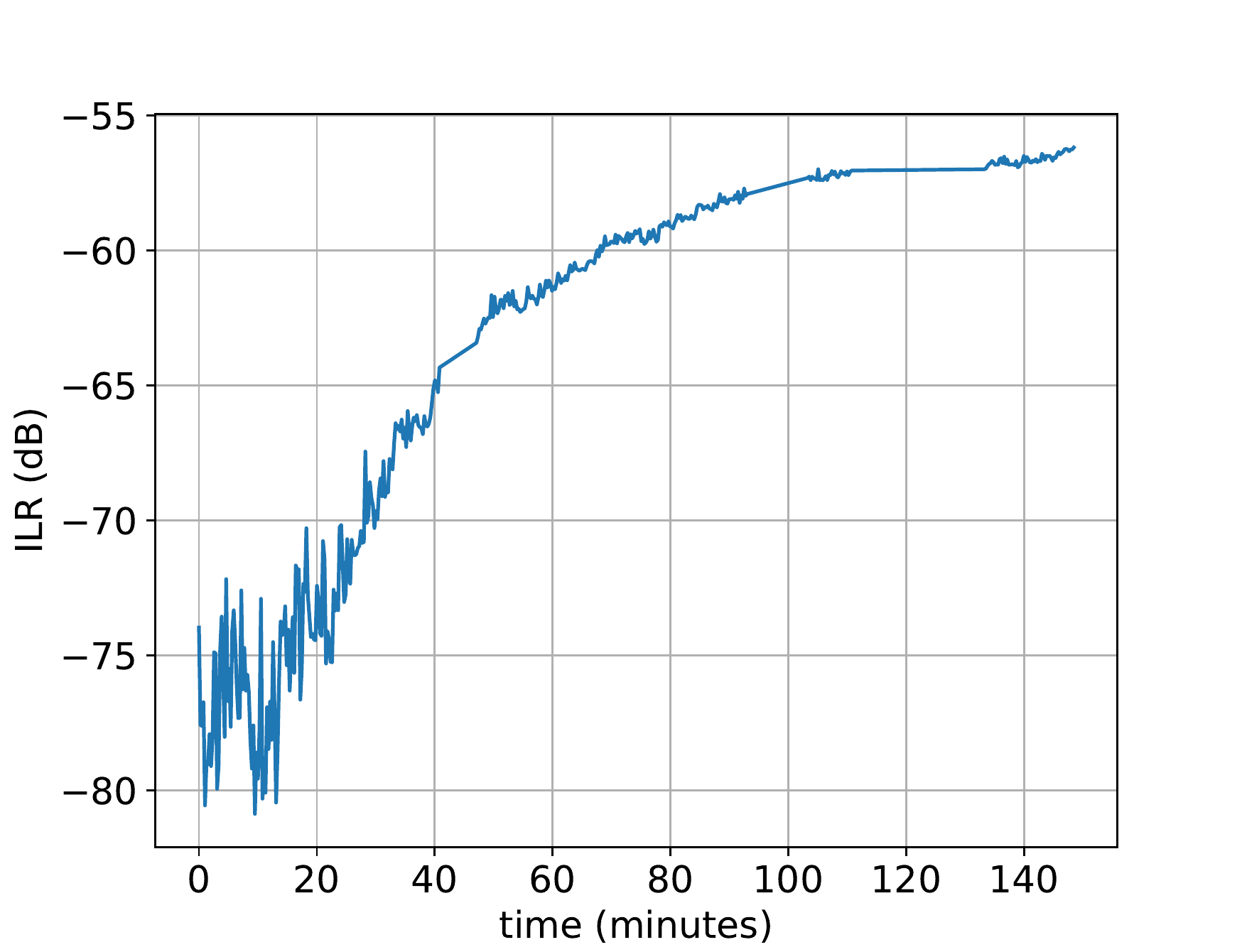}
\caption{The image leakage ratio (ILR) of an upconversion mixer as it changes over time, due to drifts in the mixer parameters. Similar phenomena will also occur at the downconversion stage. In either case, regular re-calibration is necessary for optimal performance.}
\label{fig: drift_example}
\end{figure}

The ideal calibration method should avoid any additional external equipment such as RF sources and spectrum analyzers, and it should calibrate both mixers simultaneously.
In addition it should also permit frequent re-calibration of the mixers to compensate for drift, which can be substantial. For example, in Fig. \ref{fig: drift_example} we plot the image leakage ratio (ILR), defined as the ratio of power at the image band to the power at the signal band, vs. time.
Initially the image is suppressed, but gradually increases over time by \SI{20}{dB}.
We attribute this drift to changes in the mixer's temperature.
This strongly suggests that regular re-calibration of mixers is necessary in order to maintain optimal performance of superconducting qubit processors.
For frequent re-calibration to be feasible, the both mixers must be calibrated \textit{in situ}.
In this paper, we introduce a novel technique that achieves these goals.

The paper is organized as follows: in section II we introduce the simplified IQ mixer model for upconversion and downconversion and a model for mixer imbalance. In section III we discuss the digital algorithms for calibration. In section IV we apply these techniques to a typical experimental circuit-QED setup with low-intermediate frequency (IF) transmission and receiver stage. In addition, we introduce our method to calibrate both upconversion and downconversion mixers simultaneously. In the final section V, we conclude with a brief summary.

\section{Modelling the IQ Mixer}

\begin{figure}
\includegraphics[scale=0.35]{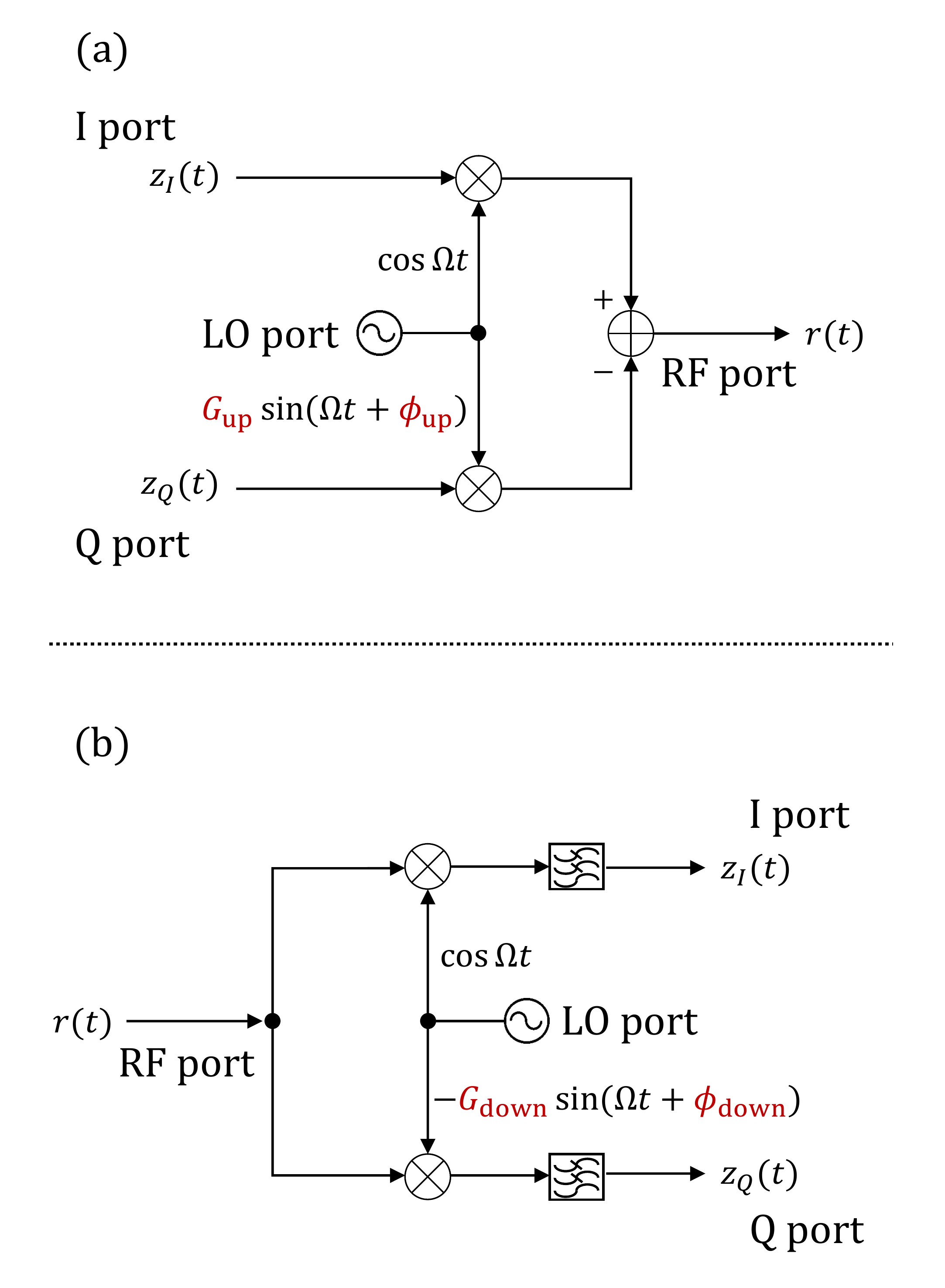}
\caption{ a) A schematic upconversion circuit with phase $\phi$ and amplitude $G$ imbalances indicated in red font.
b) A schematic downconversion circuit with imbalances indicated in red font.}
\label{fig: upconversion circuit}
\end{figure}

An ideal IQ mixer should behave as a multiplier of signals in the time domain. 
One of the signals is often a pure tone at high frequency, called a carrier wave, and the mixer is used to convert or translate the frequency components of the other tone, either up or down in frequency. 
Starting with upconversion, mixing in the time domain can be described as \cite{jKim, upconvWindisch}

\begin{align}
r(t) = \Re\{z(t)e^{j \Omega t}\},
\label{ideal_upconversion_mixer_model}
\end{align}
where $r(t)$ is the signal at the RF output port, $\Omega$ is the frequency of the carrier applied to the local-oscillator (LO) input port and $z(t) = z_I(t) + j z_Q(t)$ denotes the signals applied to the I and Q input ports (see Fig. \ref{fig: upconversion circuit} (a) ).

In the frequency domain this description is well illustrated in matrix form
\begin{align}
\begin{pmatrix}
c[\omega] \\
c^*[-\omega]
\end{pmatrix}
=\frac{1}{2}
\begin{pmatrix}
1 & 0 \\
0 & 1
\end{pmatrix}
\begin{pmatrix}
z[\Omega + \omega] \\
z^*[\Omega - \omega]
\end{pmatrix},
\label{ideal_upconversion_FT}
\end{align}
where $r[\omega] = c[\omega] + c^*[-\omega]$.
Upconversion mixing will therefore create two sidebands at $\Omega \pm \omega$.
In this paper, we refer to the upper sideband $\Omega + \omega$ as the signal and the lower sideband $\Omega - \omega$ as the image.

The image is generally undesirable as it crowds the limited bandwidth, otherwise needed for multiplexing, with unwanted signals.
This frequency crowding is particularly detrimental to applications in wireless local-area networks (WLAN) \citep{Gu2018, dirtyRF}, as well to the fidelity of frequency-multiplexed control and readout of superconducting qubits\citep{MultiplexedReadout, Kundu2019}.
The image can be cancelled by injecting the appropriate signals into the I and Q ports of the IQ mixer, such that $z_I(t) + j z_Q(t) = x(t)e^{j\omega t}$, where $x(t)$ is real.
 
However, this cancellation is negatively affected by imperfections in the mixer.
We will characterize these imperfections as amplitude $G_{\text{up}}$ and phase $\phi_{\text{up}}$ imbalance, stemming from an asymmetry in the I and Q branches of the mixer.
The imbalance modifies Eq. \eqref{ideal_upconversion_mixer_model} according to
\begin{align}
r(t) = \Re\{z(t)\left[\cos \left(\Omega t \right) + j G_{\text{up}} \sin \left(\Omega t + \phi_{\text{up}}\right)\right]\},
\label{nonideal_upconversion_mixer_model}
\end{align}
thus Eq. \eqref{ideal_upconversion_FT} becomes
\begin{align}
\begin{pmatrix}
c[\omega] \\
c^*[-\omega]
\end{pmatrix}
=\frac{1}{4}
\begin{pmatrix}
{J^*_{\text{up}}} & {K_{\text{up}}} \\
{K^*_{\text{up}}} & {J_{\text{up}}}
\end{pmatrix}
\begin{pmatrix}
z[\Omega + \omega] \\
z^*[\Omega - \omega]
\end{pmatrix},
\label{nonideal_upconversion_FT}
\end{align}
where we define the complex quantities $J_{\text{up}} = 1 + G_{\text{up}} e^{-j\phi_{\text{up}}}$ and $K_{\text{up}} = 1 - G_{\text{up}} e^{j\phi_{\text{up}}}$.
We regain the perfect mixer relations by setting $G_{\text{up}} = 1$ and $\phi_{\text{up}} = 0$.
Due to the presence of the off-diagonal terms, injecting $z(t) = x(t)e^{j\omega t}$ will no longer cancel the image completely.
However, if the phase and amplitude imbalances were known (or equivalently, $J_{\text{up}}$ and $K_{\text{up}}$), an appropriate $z(t)$ could be generated to cancel the image.

The ideal behavior of a downconversion mixer is very similar to upconversion. In the time domain the output at the I and Q ports are given by \citep{Windisch2007, Valkama2001}
\begin{align}
z(t)= z_I(t) + j z_Q(t) = \text{LPF}\{r(t)e^{-j \Omega t}\},
\label{ideal_downconversion_mixer_model}
\end{align}
where LPF denotes a low-pass filter(see Fig. \ref{fig: upconversion circuit} (b)).
The output at the I and Q ports contain both signal and image, which we write as $z(t) = Z_\text{sig}(t) + Z_\text{img}(t)$.
In the frequency domain we have
\begin{align}
\begin{pmatrix}
Z_\text{sig}[\omega] \\
Z_\text{img}^*[-\omega]
\end{pmatrix}
=\frac{1}{2}
\begin{pmatrix}
1 & 0 \\
0 & 1
\end{pmatrix}
\begin{pmatrix}
Y_\text{sig}[\omega + \Omega] \\
Y_\text{img}^*[-\omega + \Omega]
\end{pmatrix}.
\label{ideal_downconversion_FT}
\end{align}
where $Y_\text{sig}[\omega + \Omega]$ and $Y_\text{img}[\omega + \Omega]$ is the signal and image respectively entering the RF input port.

Amplitude and phase imbalance are introduced in an analogous manner. The time domain expression is modified according to
\begin{align}
z(t) = \text{LPF}\{r(t)\left[\cos \left(\Omega t \right) - j G_{\text{down}} \sin \left(\Omega t + \phi_{\text{down}}\right)\right]\},
\label{nonideal_downconversion_mixer_model}
\end{align}
and the analogous frequency-domain expression is \citep{dirtyRF, Windisch2007, Valkama2001, Windisch2005, ValkamaArticle2001}
\begin{align}
\begin{pmatrix}
Z_\text{sig}[\omega] \\
Z_\text{img}^*[-\omega]
\end{pmatrix}
=\frac{1}{4}
\begin{pmatrix}
J_{\text{down}} & K_{\text{down}} \\
{K^*_{\text{down}}} & {J^*_{\text{down}}}
\end{pmatrix}
\begin{pmatrix}
Y_\text{sig}[\omega + \Omega] \\
Y_\text{img}^*[-\omega + \Omega]
\end{pmatrix}.
\label{nonideal_downconversion_FT}
\end{align}
The definitions for $J_{\text{down}}$ and $K_{\text{down}}$ are analogous to the upconversion case.
Since the matrix is no longer diagonal, imbalance will cause downconversion from both signal and image sidebands of the carrier.

Further simplification is made by decomposing the matrix in Eq. \eqref{nonideal_downconversion_FT} into \citep{Windisch2007}
\begin{align}
\begin{pmatrix}
Z_\text{sig}[\omega] \\
Z_\text{img}^*[-\omega]
\end{pmatrix}
= \frac{1}{2}
\underbrace{
\begin{pmatrix}
1 & k_q \\
k_q^* & 1
\end{pmatrix}
}_\text{leakage matrix}
\underbrace{
\begin{pmatrix}
J_{\text{down}} & 0 \\
0 & J^*_{\text{down}}
\end{pmatrix}
}_\text{scaling matrix}
\begin{pmatrix}
Y_\text{sig}[\omega + \Omega] \\
Y_\text{img}^*[-\omega + \Omega]
\end{pmatrix},
\label{nonideal_downconversion_double_matrix}
\end{align}
where $k_q$ is the ratio 
\begin{align}
k_q =  \frac{K_{\text{down}}}{J^*_{\text{down}}} = \frac{1-G_{\text{down}}e^{j\phi_{\text{down}}}}{1+G_{\text{down}} e^{j \phi_{\text{down}}}}.
\end{align}

\section{Correcting the IQ Mixer}

In this section we take a look at how to calibrate the IQ mixers at both the upconversion and downconversion stage. We start by discussing the  upconversion mixer calibration in subsection \ref{upconversion_calibration_section} and then proceed to discuss the downconversion calibration in subsection \ref{downconversion_calibration_section}.

\subsection{Upconversion stage} \label{upconversion_calibration_section}

For upconversion, the signals $z_I$ and $z_Q$ are typically generated with a digital source, allowing for fine tuning of their amplitude and phase so as to cancel the image. We refer to this process as digital pre-distortion. The quality of the cancellation is measured by the image leakage ratio (ILR) defined as,
\begin{align}
\text{ILR} = \frac{\text{power at image frequency}}{\text{power at signal frequency}}.
\label{eqn: ILR def}
\end{align}

\begin{figure}
\includegraphics[scale=0.25]{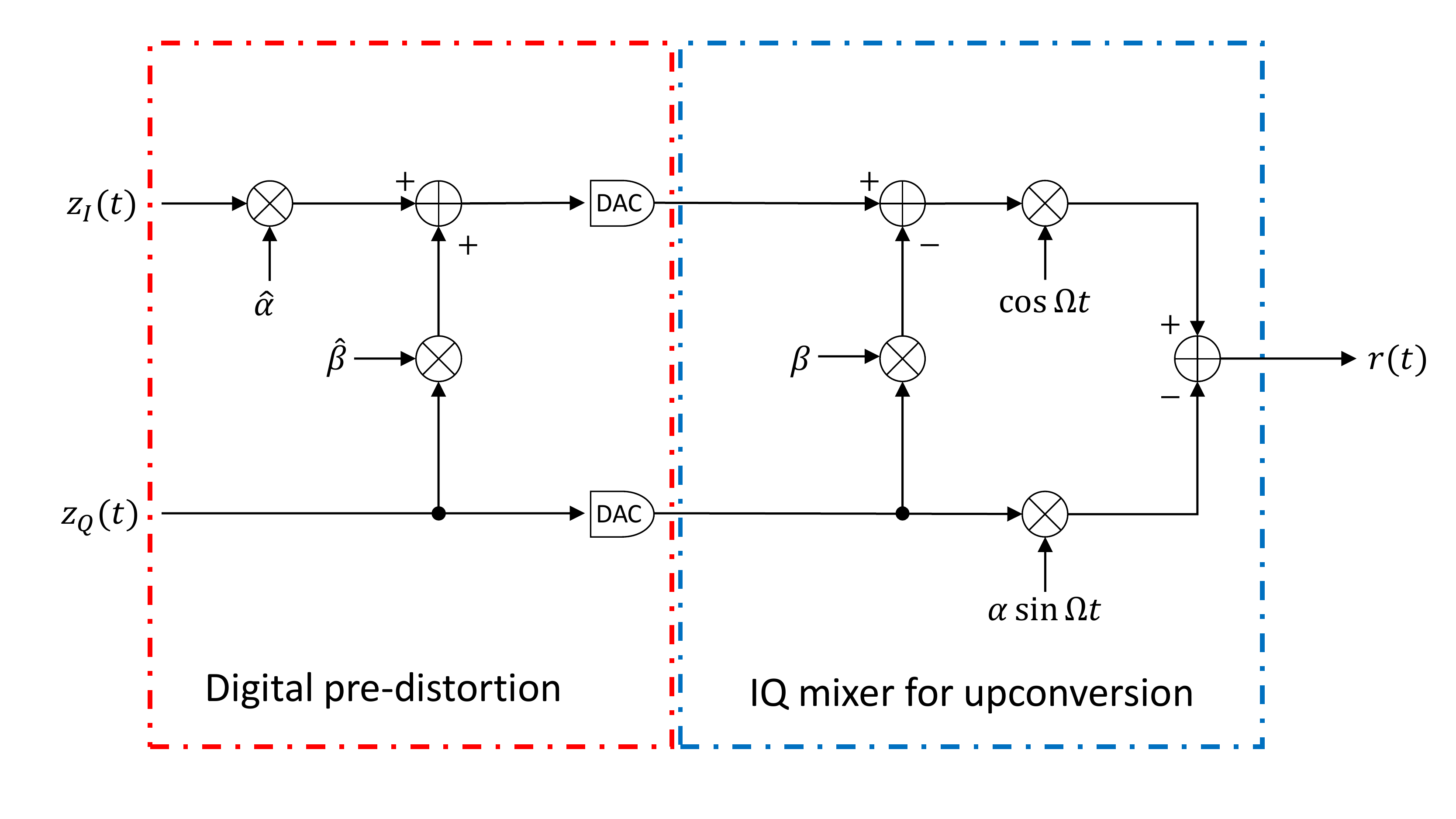}
\caption{Schematic for the upconversion circuit. The amplitude and phase imbalances $\alpha$ and $\beta$ are compensated by digitally distorting the IF input as indicated in the figure. See Ref. \citep{upconvWindisch} for details.}
\label{fig: upconversion_model}
\end{figure}

To correct for this drift one can introduce feedback to adjust the amplitude and phase of the digital pre-distortion signals at the I and Q ports, so as to continuously minimize the ILR.
The upconversion circuit schematic in Fig. \ref{fig: upconversion circuit} (a) is extended to include the digital source, depicted in Fig. \ref{fig: upconversion_model}. 
The amplitude and phase imbalances of the upconversion mixer are replaced by the alternative imbalance parameters $\alpha = G_{\text{up}} \cos \phi_{\text{up}}$ and $\beta = G_{\text{up}} \sin \phi_{\text{up}}$.
In the digital pre-distortion stage these parameters are denoted as $\hat{\alpha}$ and $\hat{\beta}$.
The ILR can then be expressed as \citep{upconvWindisch}
\begin{align}
\text{ILR}(\hat{\alpha}, \hat{\beta}) = \frac{(\alpha - \hat{\alpha})^2 + (\beta - \hat{\beta})^2}{(\alpha + \hat{\alpha})^2+ (\beta - \hat{\beta})^2}, \label{eqn: ILR}
\end{align}
revealing that when $\hat{\alpha} = \alpha$ and $\hat{\beta} = \beta$ we have optimal image suppression. 
In the absence of any digital pre-distortion, i.e. $\hat{\alpha} = 1$ and $\hat{\beta} = 0$, the ILR then depends on the intrinsic imbalances of the mixer, which vanishes for the perfectly balanced mixer $\alpha = 1$ and $\beta = 0$.

We can use semi-analytical methods \citep{jKim} or use various minimization techniques such as the Nelder-Mead method \citep{Ankel} to identify parameters $\hat{\alpha}$ and $\hat{\beta}$.
In this work we apply an iterative algorithm from Ref. \citep{upconvWindisch}.

The cost function to minimize is the quantity:
\begin{equation}
\tilde{C}(\hat{\alpha}_i,\hat{\beta}_j) = 4 \hat{\alpha}_i^2 \text{ILR}_{i, j} \label{eqn: C}
\end{equation}
where the subscripts $i$ and $j$ track the iteration steps and $\text{ILR}_{i,j}$ is the measured image leakage ratio with the applied pre-distortion parameters $\hat{\alpha}_i$ and $\hat{\beta}_j$.

The updated distortion parameter $\hat{\alpha}_i$ is calculated according to 
\begin{align}
\hat{\alpha}_i &= \frac{1}{2} \left( \hat{\alpha}_{i - 2} + \hat{\alpha}_{i - 1} - \frac{ \tilde{C}(\hat{\alpha}_{i-2},\hat{\beta}_{i-2}) - \tilde{C}(\hat{\alpha}_{i-1},\hat{\beta}_{i-2}) }{\hat{\alpha}_{i - 2} - \hat{\alpha}_{i - 1}} \right), \label{eqn: update_alpha_rule}
\end{align}
while the updated $\hat{\beta}_i$ is calculated according to
\begin{align}
\hat{\beta}_i &= \frac{1}{2} \left( \hat{\beta}_{i - 2} + \hat{\beta}_{i - 1} - \frac{ \tilde{C}(\hat{\alpha}_{i-1},\hat{\beta}_{i-2}) - \tilde{C}(\hat{\alpha}_{i-1},\hat{\beta}_{i-2}) }{\hat{\beta}_{i - 2} - \hat{\beta}_{i - 1}} \right). \label{eqn: update_beta_rule}
\end{align}

We initialize the algorithm by choosing $\hat{\alpha}_0, \hat{\alpha}_1 \approx 1$, $\hat{\beta}_0 , \hat{\beta}_1 \approx 0$, and proceed as follows:

\begin{enumerate}
\item Distort signals by $\hat{\alpha}_0$ and $\hat{\beta}_0$, measure $\tilde{C}(\hat{\alpha}_0,\hat{\beta}_0)$.

\item Distort signals by $\hat{\alpha}_1$ and $\hat{\beta}_0$, measure $\tilde{C}(\hat{\alpha}_1,\hat{\beta}_0)$.

\item Distort signals by $\hat{\alpha}_1$ and $\hat{\beta}_1$, measure $\tilde{C}(\hat{\alpha}_1,\hat{\beta}_1)$. Compute $\hat{\alpha}_2$.

\item Distort signals by $\hat{\alpha}_2$ and $\hat{\beta}_1$, measure $\tilde{C}(\hat{\alpha}_2,\hat{\beta}_1)$. Compute $\hat{\beta}_2$.

\item Distort signals by $\hat{\alpha}_2$ and $\hat{\beta}_2$, measure $\tilde{C}(\hat{\alpha}_2,\hat{\beta}_2)$. Compute $\hat{\alpha}_3$.

\item Distort signals by $\hat{\alpha}_3$ and $\hat{\beta}_2$, measure $\tilde{C}(\hat{\alpha}_3,\hat{\beta}_2)$. Compute $\hat{\beta}_3$.
\\
\centerline{\vdots}

\end{enumerate}
The iteration is continued until a desired threshold on the $\text{ILR}_{i, j}$ is satisfied.

\subsection{Downconversion stage} \label{downconversion_calibration_section}

Given an observed signal $Z_\text{sig}[\omega]$ and idler $Z^*_\text{img}[-\omega]$ after downconversion, we want to reconstruct the actual signal $Y_\text{sig}[\omega + \Omega]$ and idler $Y^*_\text{img}[-\omega + \Omega]$ before entering the imbalanced IQ mixer. 
This is possible according to Eq. \eqref{nonideal_downconversion_double_matrix}, if we manage to determine $k_q$ and $J_{\text{down}}$.
For simplicity, the scaling matrix is henceforth neglected since it does not contribute to the leakage between sidebands.
This leaves us with only one complex parameter $k_q$.

Before proceeding, we simplify our notation by introducing $Y_m = Y_\text{sig}[\omega + \Omega]$ and $Y^*_{-m} = Y^*_\text{img}[-\omega + \Omega]$. 
The subscript $-m$ denotes the images while $m$ denotes signals. 
Analogous notation will also be used for $Z_\text{sig}[\omega]$ and $Z^*_\text{img}[-\omega]$.
The current best estimate of $k_q$ is denoted $\hat{k}_q$.
With $\hat{k}_q$ we reconstruct the undistorted signal and image according to
\begin{align}
\begin{pmatrix}
Y_m \\
Y^*_{-m}
\end{pmatrix}
=
\begin{pmatrix}
1 & -\hat{k}_q \\
-\hat{k}^*_q & 1
\end{pmatrix}
\begin{pmatrix}
Z_m \\
Z^*_{-m}
\end{pmatrix}, \label{mixing_model}
\end{align}
neglecting the complex scaling factor.

We use a method developed by Windisch and Fettweis \citep{dirtyRF, Windisch2005, WindischKalman2007} to estimate $\hat{k}_q$. If we assume $\langle Y_m Y_{-m} \rangle = 0$, it is possible to estimate the mixer imbalance by first calculating:
\begin{align}
\hat{k}_p &=  \frac{\left\langle{Z_m Z_{-m}}\right\rangle}{\left\langle\left|Z_m + Z_{-m}^* \right|^2 \right\rangle}, \label{downconversion_calibration_eq1}
\end{align}
where $\langle ... \rangle$ indicates an average.
Then $\hat{k}_q$ is related to $\hat{k}_p$ according to:
\begin{align}
\hat{k}_q &= \frac{1 - a - j b}{1+a+j b},\\
b &= -2 \Im \left[ \hat{k}_p \right], \\
a &= \sqrt{1-b^2-4 \Re\left\{ \hat{k}_p \right\}},
\label{downconversion_calibration_eq5}
\end{align}
Consequentially, the amplitude and phase imbalances are $G_{\text{down}} = \sqrt{a^2 + b^2}$ and $\phi_{\text{down}} = \arctan(b/a)$ respectively. 
Equations \eqref{downconversion_calibration_eq1} - \eqref{downconversion_calibration_eq5} are the core of the downconversion calibration procedure. 
Because the downconversion mixer also drifts in time (see Fig. \ref{fig: drift_example}), feedback is also needed to update $\hat{k}_q$.  The procedure is therefore enhanced with a Kalman filter.

Kalman filters are used to determine state variables/parameters (stationary or dynamical) from noisy observations. The filter is employed in such diverse applications as navigation \citep{KalmanBook} and multifrequency atomic force-microscopy \citep{Harcombe2020}.
Other applications include cooling of nanoparticles through feedback \citep{Setter2018}, and state estimation in atomic magnetometry \citep{Geremia2003} and optomechanics \citep{Witlef2015, Chen2013}. 
For a detailed theoretical discussion, we refer the interested reader to Ref. \citep{Windisch2007, WindischKalman2007}.
To set up a discrete-time Kalman filter, we need a process equation and a measurement equation \citep{KalmanBook}.
 
The process equation relates the sought after state variables at two different points in time. 
In this case, the state variable is $k_q$ and our process equation is 
\begin{align}
k_q(i+1) &= k_q(i) + v_p(i). \label{process_noise}
\end{align}
The argument $i$ denotes ordered, finite intervals of time that we henceforth refer to as time frames.
The IQ mixer state variable $k_q(i)$ is assumed to evolve in time according to equation \eqref{process_noise},  in which the process noise $v_p(i)$ is Gaussian with zero mean and variance $\sigma^2_p (i)$.
We attribute the origin of this noise term to various fluctuations in the analog electronics and temperature.

The measurement equation describes how our estimated value of the state variable, which is determined from noisy measurement data, is related to the actual or true value of the state variable
\begin{align}
\hat{k}_q(i) &= k_q(i) + v_m(i). \label{measaurement_noise}
\end{align}
The measurement noise $v_m(i)$ is Gaussian with zero mean and variance $\sigma^2_m (i)$.

The Kalman filter provides the estimated $\hat{k}_q$ at different time frames.
The filtered estimate $\hat{k}_q(i|\mathfrak{H}_{i})$, with variance $\sigma_m^2(i)$, is the best estimate (in the least-square sense) of the state $k_q$ at the current time frame $i$, based on a history of observations $\mathfrak{H}_{i}$ up to and including frame $i$.
The predicted estimate $\hat{k}_q(i+1|\mathfrak{H}_{i})$, with variance $\sigma^2_m(i+1, i)$, is the best estimate for the future frame $i+1$ based on the past history of observations.

The Kalman filter is initialized at $i=0$ such that $\hat{k}_q(1|\mathfrak{H}_{0}) = 0$ and its variance $\sigma^2_m(1,0) = \infty$.
Other values of $\hat{k}_q(1|\mathfrak{H}_{0})$ with smaller $\sigma^2_m(1,0)$ are possible (for example, if we have values from a previous calibration run).
For $i \geqslant 1$ we do the following \citep{Windisch2007}:
\begin{enumerate} 
\item Record $n=1,2, ..., N$ samples of $Z^n_k(i)$, $Z^n_{-k}(i)$ and reconstruct $Y^n_k(i)$, $Y^n_{-k}(i)$ using the predicted $\hat{k}_q(i|\mathfrak{H}_{i-1})$ according to Eq. \eqref{mixing_model}. Thereafter, estimate the variances, $\sigma^2_k(i)$ and $\sigma^2_{-k}(i)$ from the $Y^n_k(i)$ and $Y^n_{-k}(i)$ data.
\item Simultaneously, use $Z^n_k(i)$, $Z^n_{-k}(i)$ to compute $\hat{k}_q(i)$ according to Eq. \eqref{downconversion_calibration_eq1}. Here $\hat{k}_q(i)$ is the current time frame's noisy estimate. The error variance of this quantity is approximately
\begin{align}
\sigma_{q}^2(i) &\approx \left[N \left(1 + \frac{\sigma^2_{k}}{\sigma^2_{-k}} \right) \left(1 + \frac{\sigma^2_{-k}}{\sigma^2_{k}} \right) \right]^{-1}.
\end{align}
\item Update the variance of the filtered estimate as
\begin{align}
\frac{1}{\sigma_m^2(i)} &= \frac{1}{\sigma_m ^2(i, i-1)}+ \frac{1}{\sigma_{q}^2(i)}.
\end{align}
\item Update the predicted estimate 
\begin{align}
\hat{k}_q(i+1|\mathfrak{H}_{i}) &= \sigma_m^2(i) \left( \frac{\hat{k}_q(i|\mathfrak{H}_{i-1})}{\sigma_m ^2(i, i-1)} + \frac{\hat{k}_q(i)}{\sigma_{q}^2(i)} \right).
\end{align}
\item Compute the variance of the predicted estimate as
\begin{align}
\sigma_m^2(i+1, i) &= \sigma_m^2(i) + \sigma_p^2(i).
\end{align}
\item Finally, update the filtered estimate
\begin{align}
\hat{k}_q(i|\mathfrak{H}_{i}) &= \hat{k}_q(i+1|\mathfrak{H}_{i}).
\end{align}

\end{enumerate}
Notice that the drift $\sigma_p^2(i)$ is not updated.
This needs to be specified from the start and its magnitude (compared to $\sigma_m^2(i)$) will determine how little weight is given to past in contrast to present observations.

\begin{figure}
\includegraphics[scale=0.5]{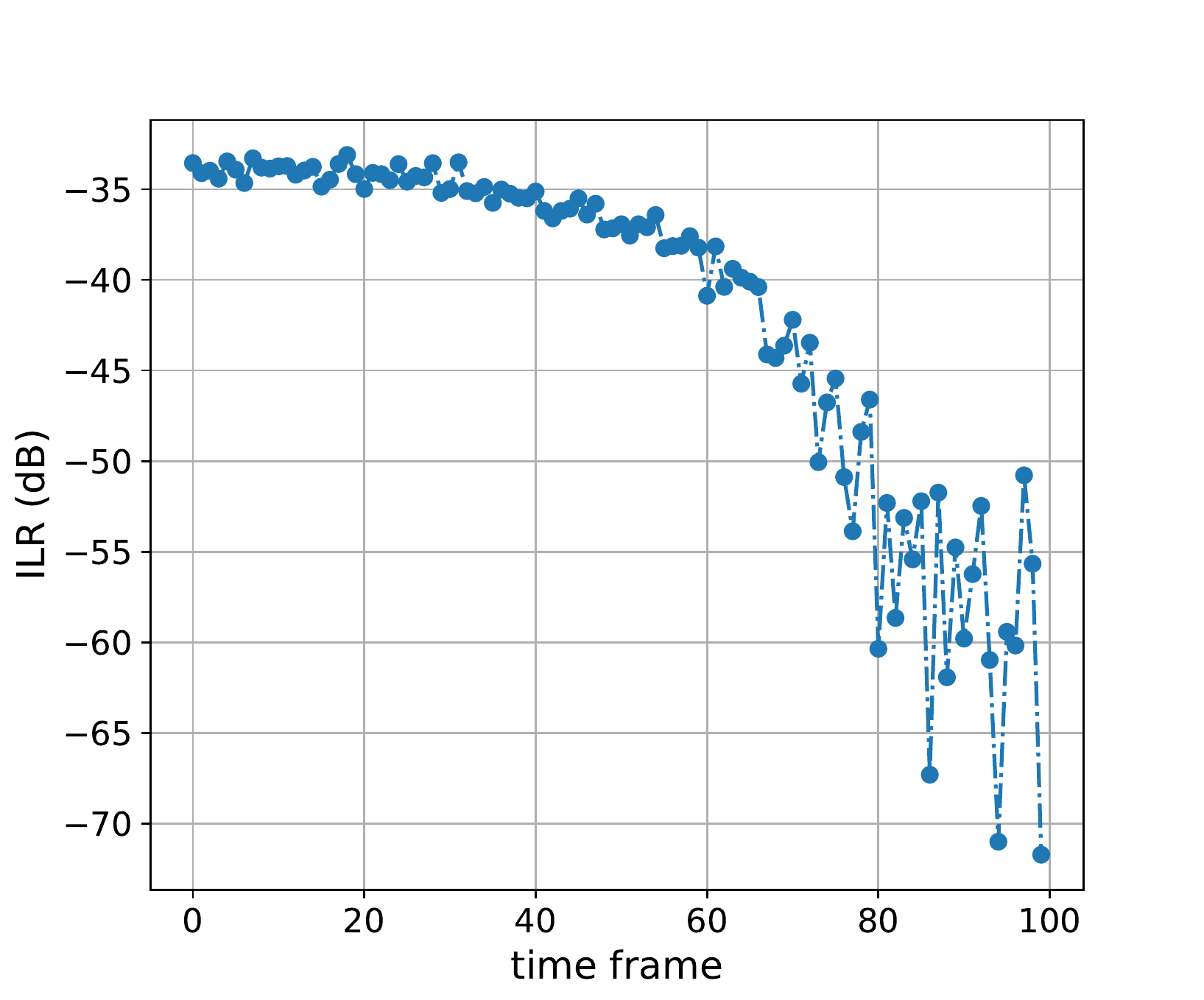}
\caption{Reconstructed ILR through the downconversion mixer using a Kalman filter as a function of time frames.
We initialize the algorithm by assuming a perfect mixer $G_{\text{down}} = 1$ and $\phi_{\text{down}} = 0$, with near-zero variance.
After 100 iterations, the ILR becomes limited by the noise floor and we extract the filtered parameters $G_{\text{down}} = 0.961$ and $\phi_{\text{down}} = 0.96^\circ$.}
\label{fig: pure USB tone calibration}
\end{figure}

We test the filter on a system consisting only of a downconversion mixer driven by a pure RF tone corresponding to the signal.
The Kalman filter is initialized with the erroneous assumption that the mixer is ideal with near-zero variance.
The incoming RF signal is integrated for 4 ms and the Kalman filter algorithm is used to acquire an estimate of the filtered parameter $\hat{k}_q(i|\mathfrak{H}_i)$.
The image and signal sidebands are reconstructed according to Eq. \eqref{mixing_model}, whose quality can be assessed by plotting the ILR at each iteration step, see Fig. \ref{fig: pure USB tone calibration}.
There is a steady decrease in the ILR until the noise floor is reached, indicating a successful calibration of the downconversion mixer.

\begin{figure}
\includegraphics[scale=0.5]{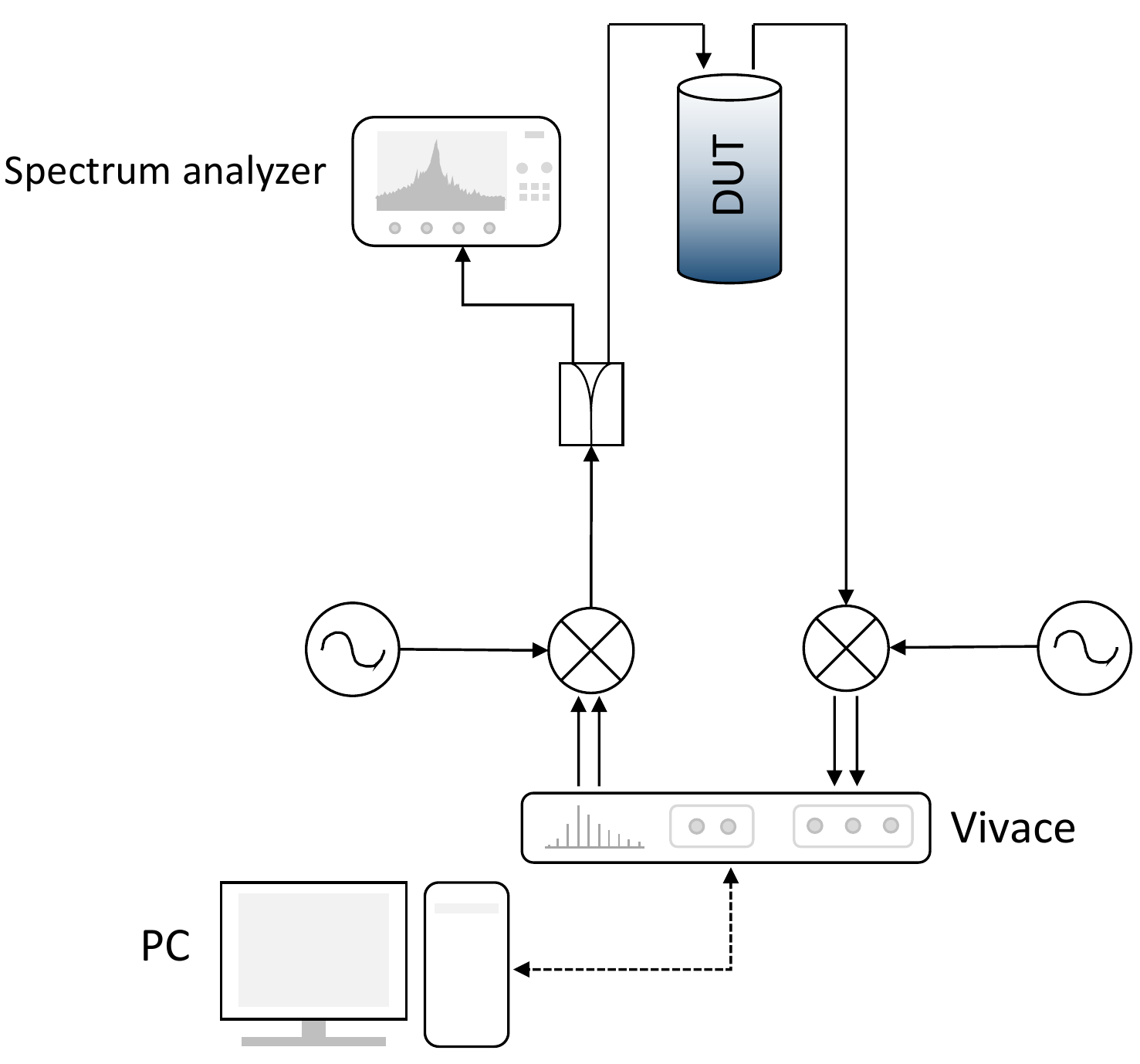}
\caption{Schematic setup for mixer calibration. The output RF signal is split with a Wilkinson power divider. One branch leads to a fully equipped cryogenic low-noise microwave reflection setup inside a dilution refrigerator.
The other branch is monitored by a spectrum analyzer, to prove our procedure is working as intended.}
\label{fig:set_up}
\end{figure}

\section{Complete calibration of conversion chain}

\begin{figure*}
\includegraphics[scale=0.7]{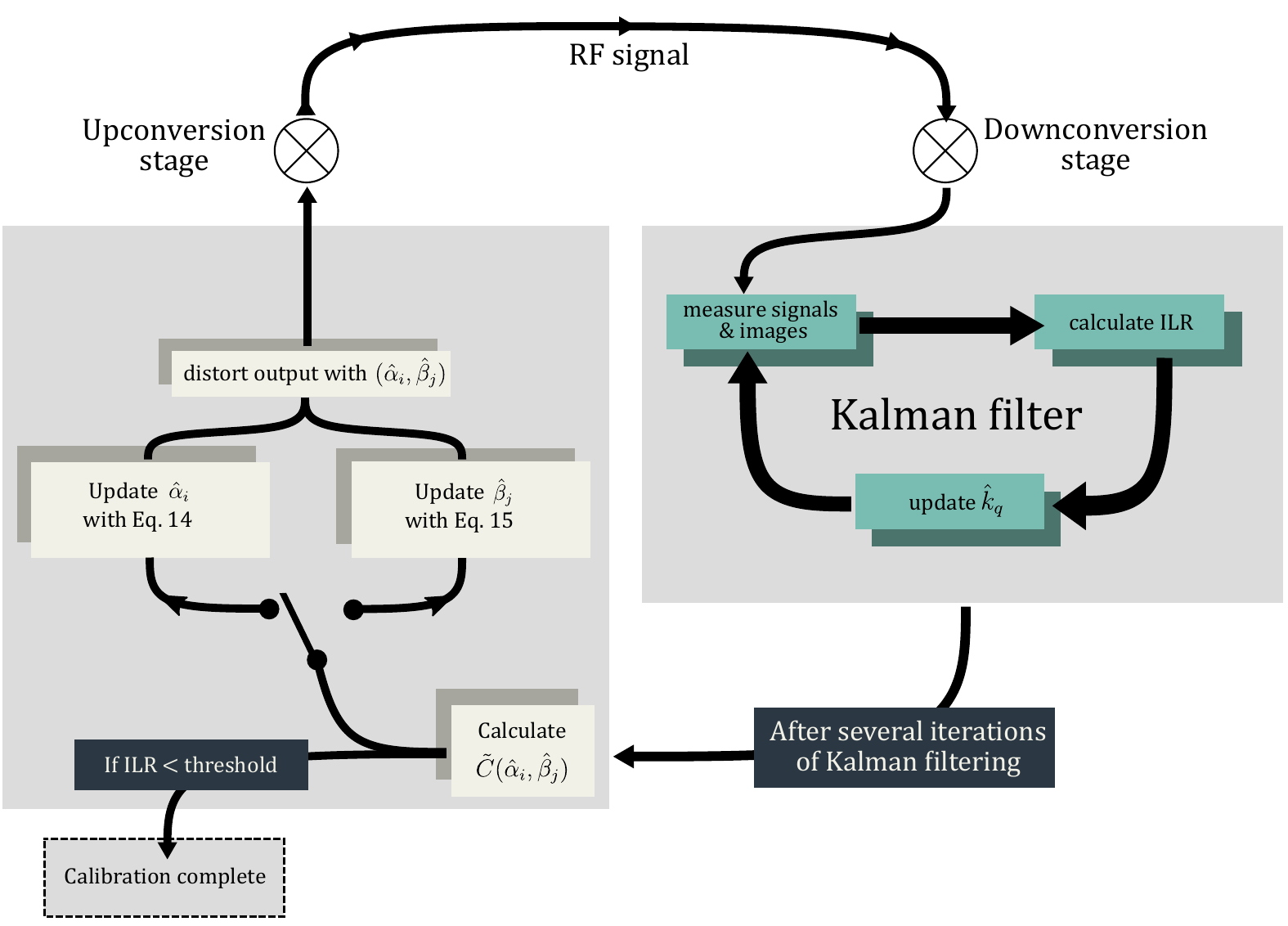}
\caption{Flow chart graphically summarises the calibration loop.
The upconversion calibration alternately updates $\alpha$ and $\beta$ using an estimated value of $\tilde{C}(n, m)$ obtained from the downconversion stage.
A Kalman filter is used to calibrate the downconversion stage to ensure a correct value of $\tilde{C}(n, m)$ is obtained.
}
\label{fig:flow_chart}
\end{figure*}

The upconversion and downconversion mixers are individually straightforward to calibrate, as previously shown. However, many experimental setups consists of both upconversion and downconversion stages, each with imbalances that needs to be corrected. 
Calibrating both mixers \textit{in situ} introduces another problem not addressed so far: how does one go about distinguishing the error of one mixer from another? 
We introduce and demonstrate two new approaches that resolves this issue and permits automated calibration of the entire frequency conversion chain.

These ideas are tested on a fairly typical experimental setup for research in circuit QED, shown in Fig. \ref{fig:set_up}.
The digital source used to generate the I and Q signals is the Vivace platform from Intermodulation Products AB \citep{Vivace}. 
This same instrument also coherently demodulates the downconverted signals. 
With multiple output and input ports to drive and measure at the I and Q ports of both up- and downconversion mixers, and with its ability to coherently modulate and demodulate up to 32 frequencies simultaneously, Vivace greatly simplifies our measurement setup.
The output is upconverted with a IQ0318M mixer from Marki Microwave and split with a power divider into two branches.
One branch is connected to a spectrum analyzer.
The spectrum analyzer is not required and its sole purpose is to act as an independent observer of the calibration procedure.
Meanwhile, the other branch is connected to a dilution refrigerator, wired with low-noise amplifiers, attenuators, circulators and isolators.
The signal is finally downconverted with another IQ0318M mixer, before being digitally demodulated from intermediate frequency (IF) to baseband with Vivace.

The ILR produced by the imbalanced upconversion mixer can be measured by both the spectrum analyzer and the imbalanced downconversion stage.
If the downconversion calibration works ideally, the ILR measured at the downconversion stage (henceforth referred to as the downconversion ILR) will be identical to the ILR measured with the spectrum analyzer (referred to as the upconversion ILR). 

The calibration loop is graphically summarized in Fig. \ref{fig:flow_chart}. 
It can be divided into two sections, corresponding to the up- and downconversion stages. 
The loop principally follows the algorithm in section \ref{upconversion_calibration_section}. 
The new addition is the Kalman filter (see section \ref{downconversion_calibration_section}) at the downconversion stage.  
The Kalman filter is run for several iterations (in our case, the number of iterations ranged from 20-100) to calibrate the downconversion mixer, which is necessary to acquire an accurate value for $\tilde{C}(\hat{\alpha}_i,\hat{\beta}_j)$ (see Eq. \eqref{eqn: C}).
This in turn suggests the next $\hat{\alpha}_i$ or $\hat{\beta}_i$ through Eq. \eqref{eqn: update_alpha_rule} and \eqref{eqn: update_beta_rule}.
However, the imbalanced upconversion-mixer will generate correlated signals and images, violating the assumption $\langle Y_m Y_{-m} \rangle = 0$ necessary for downconversion-mixer calibration. 

We present two methods to circumvent this problem.
The first method involves calibrating with white noise at the signal and image bands. 
While calibrating the downconversion mixer, we turn off the upconversion-mixer output, and sample noise instead. 
For the sake of clarity, we explicitly list the first few steps of the procedure. First initialize by choosing $\hat{\alpha}_1$ and $\hat{\alpha}_2$ close to 1, $\hat{\beta}_1$ and $\hat{\beta}_2$ close to 0, then:

\begin{enumerate}
\item Turn \textit{off} the RF upconversion-mixer output (by turning off the IF signal).

\item Run the Kalman filter (according to section \ref{downconversion_calibration_section}) by sampling incoming noise in the signal and image bands. This results in a best $\hat{k}_{q}$.

\item Turn \textit{on} the RF upconversion-mixer output. Distort signals by $\hat{\alpha}_1$ and $\hat{\beta}_1$. Measure both signal and image sidebands at the downconversion stage. 

\item Proceed to reconstruct the undistorted sidebands according to Eq. \eqref{mixing_model}, using the recently acquired $\hat{k}_{q}$. The ratio of powers is the downconversion ILR. Use it to calculate the quantity $\tilde{C}(\hat{\alpha}_1,\hat{\beta}_1)$, as defined in Eq. \eqref{eqn: C}.

\item Turn \textit{off} the RF upconversion-mixer output.

\item Run the Kalman filter (according to section \ref{downconversion_calibration_section}) by sampling incoming noise in the signal and image bands. This results in a new best $\hat{k}_{q}$.

\item Turn \textit{on} the RF upconversion-mixer output. Distort signals by $\hat{\alpha}_2$ and $\hat{\beta}_1$. Measure both signal and image sidebands at the downconversion stage. 

\item Proceed to reconstruct the undistorted sidebands according to equation \eqref{mixing_model}, using the recently updated $\hat{k}_{q}$. The ratio of powers is the downconversion ILR. Use it to calculate the quantity $\tilde{C}(\hat{\alpha}_2,\hat{\beta}_1)$.
\\
\centerline{\vdots}

\end{enumerate}
Notice how we follow the algorithm given in section \ref{upconversion_calibration_section}, with the addition of turning off the upconversion mixer output and calibrating the downconversion mixer before every new estimate of $\tilde{C}(n, m)$. The initialization of the Kalman filter would be either a best guess or using the best estimate values from the previous Kalman filter loop.

\begin{figure}
\includegraphics[scale=0.5]{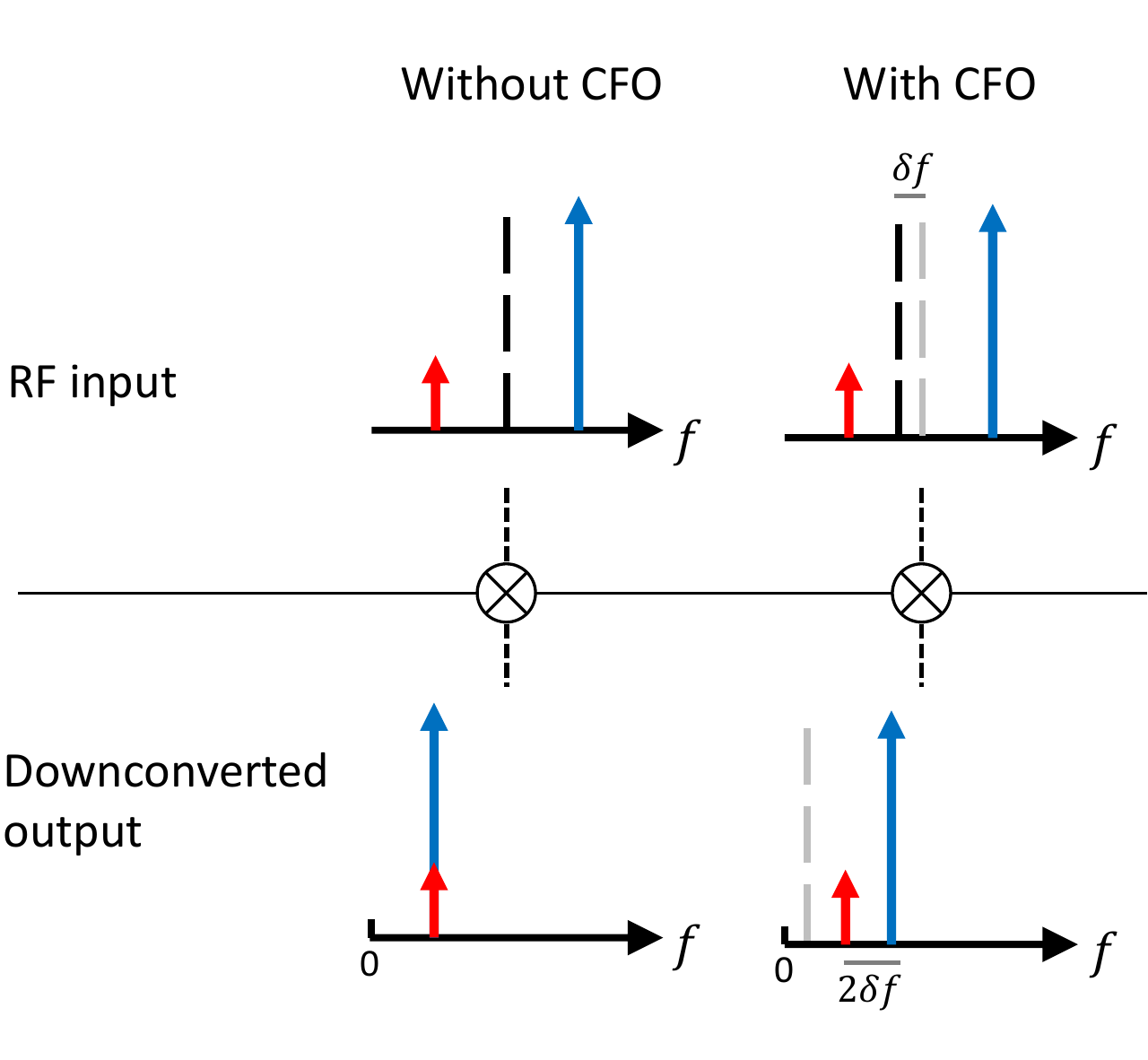}
\caption{\label{fig:CFO_schematic} Downconverting with an LO detuned by $\delta f$, referred to as a carrier frequency offset (CFO), spectrally separates the two sidebands at downconversion by $2\delta f$. As a consequence, the downconversion stage effectively measures two signal-image pairs. If we call the red and blue arrows signals, then their respective images contain only noise. Each signal-image pair is therefore uncorrelated and the downconversion calibration is valid.}
\end{figure}

The second method to separate the calibration of upconversion and downconversion relies on two different carrier tones, detuned by a carrier frequency offset (CFO). 
Referring to Fig. \ref{fig:CFO_schematic}, a CFO will prevent the incoming signal (blue) and image (red) sidebands from being superimposed upon downconversion. 
Thus from the perspective of the downconversion stage, the blue and red arrows are two incoming signals, with their respective image bands containing only noise.
The Kalman filter is run for each pair separately (essentially two filters running in parallel) and generates two values $\hat{k}^\text{red}_q$ and $\hat{k}^\text{blue}_q$.
The red and blue signals are distorted by the imbalances in the downconversion mixer, and are corrected by inserting $\hat{k}^\text{red}_q$ and $\hat{k}^\text{blue}_q$ into Eq. \eqref{mixing_model}
\begin{align}
\begin{pmatrix}
Y_\text{red} \\
Y^*_{-, \text{red}}
\end{pmatrix}
&=
\begin{pmatrix}
1 & -\hat{k}_{q, \text{red}} \\
-\hat{k}^*_{q, \text{red}} & 1
\end{pmatrix}
\begin{pmatrix}
Z_\text{red} \\
Z^*_{-, \text{red}}
\end{pmatrix},
\\
\begin{pmatrix}
Y_\text{blue} \\
Y^*_{-, \text{blue}}
\end{pmatrix}
&=
\begin{pmatrix}
1 & -\hat{k}_{q, \text{blue}} \\
-\hat{k}^*_{q, \text{blue}} & 1
\end{pmatrix}
\begin{pmatrix}
Z_\text{blue} \\
Z^*_{-, \text{blue}}
\end{pmatrix}.
\end{align}
The ILR used to calibrate the upconversion mixer is then calculated as $\text{ILR} = |Y_\text{red}|/|Y_\text{blue}|$.
Using a CFO, a proper estimation of the ILR necessitates measuring four frequency bands (two signal-image pairs), instead of the two (one signal-image pair).
The Vivace platform is well suited to this task as its multifrequency lock-in can measure up to 32 bands.
It is also worth noting that $\hat{k}^\text{red}_q$ and $\hat{k}^\text{blue}_q$ are frequency dependent and the relative difference is expected to vary as a function of CFO.
In our case, we used a CFO of 20kHz resulting in typical magnitude ratios of $|\hat{k}^\text{red}_q|/|\hat{k}^\text{blue}_q| \approx 1.007$ and phase ratios of $\angle\hat{k}^\text{red}_q/\angle\hat{k}^\text{blue}_q \approx 0.988$.

\begin{figure}
\includegraphics[scale=0.5]{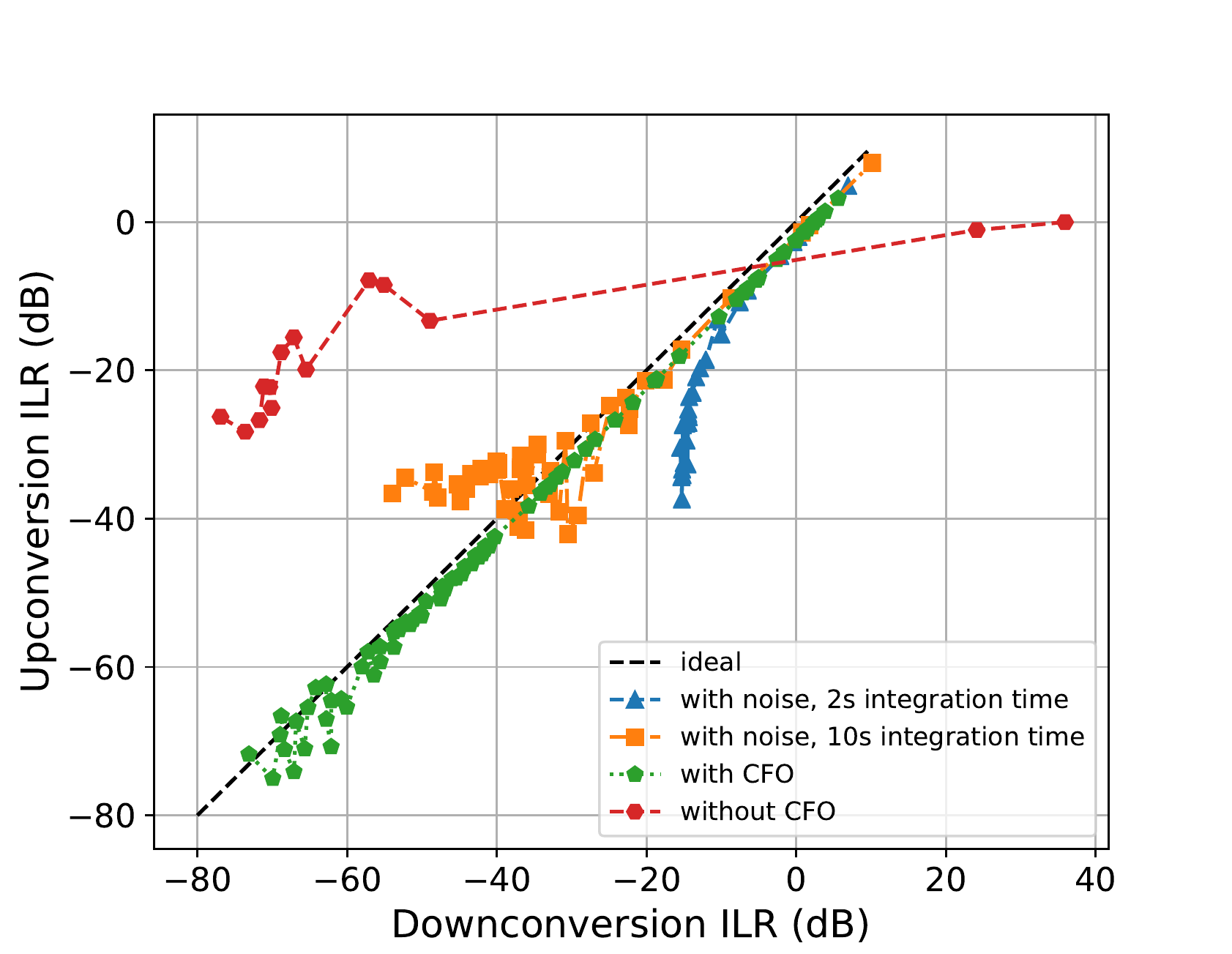}
\caption{\label{fig:up_down_conv} Comparing the upconversion and downconversion ILR. Ideally the two quantities should be equal, as indicated by the black dashed line. Major deviations are due to faulty calibration. Small deviations are attributed to frequency dependent attenuation in the cryogenic electronics. Differences between the various methods are explained in the main text.}
\end{figure}

Fig. \ref{fig:up_down_conv} plots the upconversion ILR, as measured by the spectrum analyzer, vs. the downconversion ILR, as measured by Vivace. 
Successive iterations of the calibration are plotted as points, connected with dashed lines, starting in the upper right with large ILR, and proceeding toward the lower left with reduced ILR. 
The colors represent different calibration scenarios. 
The dashed diagonal line represents the ideal case.
Notice that attempts to calibrate both up- and downconversion simultaneously while $\langle Y_m Y_{-m} \rangle \neq 0$ (red), fail to produce the ideal result, as expected.

\begin{figure}
\includegraphics[scale=0.5]{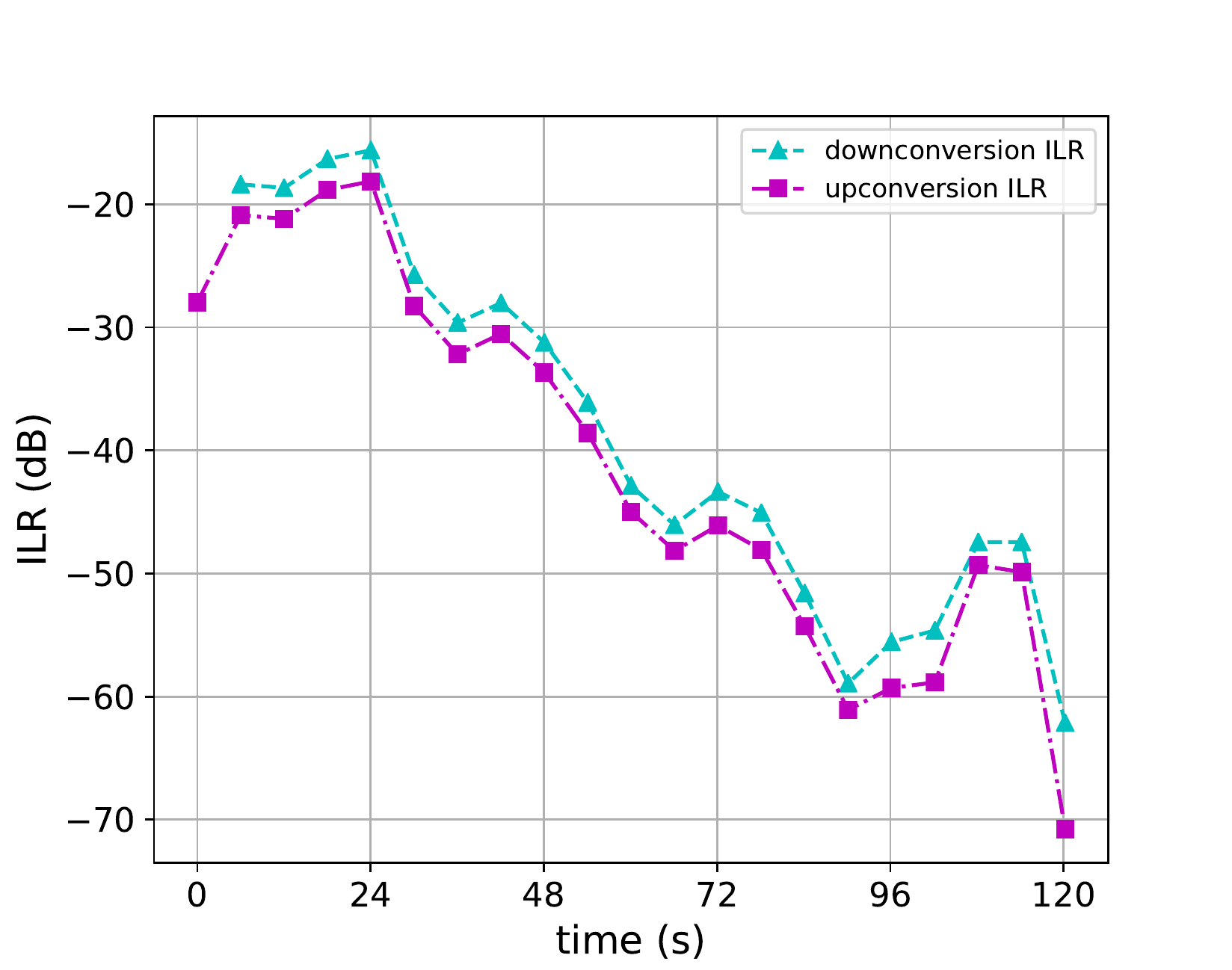}
\caption{\label{fig:CFO_down_conv} Calibration with CFO as a function of iteration number. 
Calibration time is limited by the spectrum analyzer used to monitor the upconversion ILR.
If we remove this monitor, -70 dB ILR can be reached within a couple of seconds.
The imbalance parameters are $\alpha = 0.923$ and $\beta = -0.0327$, corresponding to $G_{\text{up}} = 0.961$ and $\phi_{\text{up}} = -2.03^\circ$.}
\end{figure}

Two calibrations using only noise (blue and orange) are depicted, the difference being the amount of time spent sampling noise (2s and 10s respectively). 
The accuracy is decent for large ILR, but suffers as we approach -40 dB.

The calibration improves significantly if we use a signal and a CFO (green). 
We can accurately reconstruct the ILR down to -70 dB. 
Notice the offset of roughly 3 dB from the diagonal, corresponding to an overestimate of the ILR at the downconversion stage. 
We attribute this to an unequal attenuation of the sidebands in the cryogenic electronics.

Finally, Fig. \ref{fig:CFO_down_conv} depicts calibration with CFO of two imbalanced mixers as a function of time.
Each point corresponds to a complete loop around the chart in Fig. \ref{fig:flow_chart}. The downconversion ILR is successfully suppressed to less than -60 dB with our setup, showing that we can calibrate both up- and downconversion mixers simultaneously. The offset from the upconversion ILR is attributed to frequency-dependent attenuation in the cryogenic electronics.

\section{Conclusion}

Current IQ-mixer components for circuit QED experiments are analog devices that, due to their analog nature, come with imperfections and imbalances.
These imbalances can be mitigated by digital pre-distortion techniques. 
In this paper, we have presented several such techniques to calibrate IQ mixers, both for up- and downconversion. 
These techniques, originally developed for OFDM applications, are shown to be suitable for circuit-QED applications. 
By careful combination of up- and downconversion techniques, it is possible to calibrate both stages simultaneously. 
In addition, we show that it is possible to calibrate the mixers \textit{in situ}, provided the frequency-dependent attenuation is not too large. 
We believe this will pave the way for more efficient mixer-calibration procedures as superconducting-qubit processors continue to increase in size and complexity in the near future.

\begin{acknowledgments}
We acknowledge the financial support from the Knut and Alice Wallenberg Foundation through the Wallenberg Center for Quantum Technology (WACQT).
We also acknowledge Per Delsing and Simone Gasparinetti for helpful comments.
\end{acknowledgments}

\nocite{*}
\bibliographystyle{aipnum4-1}
\bibliography{biblio}

\end{document}